\documentclass[preprint,showpacs,aps]{revtex4}

\usepackage{graphicx}
\usepackage{dcolumn}
\usepackage{bm}

\begin{document}

\preprint{\today}

\def\d{{\partial }}
\def\al{{\alpha }}
\def\be{{\beta }}
\def\ga{{\gamma }}
\def\de{{\delta }}
\def\ep{{\epsilon }}
\def\la{{\lambda}}
\def\La{{\Lambda}}
\def\si{{\sigma}}
\def\Om{{\bf \Omega }}
\def\om{{\omega }}
\def\omc{{\omega_{c} }}
\def\t{{\tau }}
\def\T{{\cal T }}
\def\I{{\cal I}}
\def\R{{ R}}
\def\kl{{k \lambda}}
\def\ql{{q \lambda}}
\def\qo{{q_1 \lambda_1}}
\def\qt{{q_2 \lambda_2}}
\def\rto{{R_1 \tau_1}}
\def\rtt{{R_2 \tau_2}}
\def\k{{ k }}
\def\S{{\cal S}}
\def\beqnar{\begin{eqnarray}}
\def\eeqnar{\end{eqnarray}}
\def\beq{\begin{equation}}
\def\eeq{\end{equation}}
\def\bvt{{\begin{verbatim}}}
\def\evt{{\end{verbatim}}}
\def\bfig{{\begin{figure}}}
\def\efig{{\end{figure}}}
\def\bs{{\begin{split}}}
\def\es{{\end{split}}}

\title{Heat transport in silicon 
from first principles calculations}

\author{Keivan Esfarjani }
\affiliation{ Department of Mechanical Engineering, MIT, 77 Mass Av., Cambridge, MA 02139 \\}
\author{Gang Chen }
\affiliation{ Department of Mechanical Engineering, MIT, 77 Mass Av., Cambridge, MA 02139 \\}
\author{Harold T. Stokes}
\affiliation{Department of Physics, Brigham Young University,
Salt Lake City, UT 84602 \\}
\date{\today}

\begin{abstract}
Using harmonic and anharmonic force constants extracted from  
density-functional calculations within a supercell, we have 
developed a relatively simple but general method to compute 
thermodynamic and thermal properties of any crystal. 
First, from the harmonic, cubic, and quartic force constants we
construct a force field for molecular dynamics (MD). It is exact in the 
limit of small atomic displacements and thus does not suffer from
inaccuracies inherent in semi-empirical potentials such as Stillinger-Weber's.
By using the Green-Kubo (GK) formula and molecular dynamics simulations, we 
extract the bulk thermal conductivity. This method is accurate 
at high temperatures where three-phonon processes need to be included to higher orders,
but may suffer from size scaling issues.
Next, we use perturbation theory (Fermi Golden rule) to extract the 
phonon lifetimes and compute the thermal conductivity $\kappa$ from the 
relaxation time approximation.  
This method is valid at most temperatures, but will
overestimate $\kappa$ at very high temperatures, where higher order processes 
neglected in our calculations, also contribute.
As a test, these methods are applied to bulk crystalline silicon,
and the results are compared and differences discussed in more detail.
The presented methodology paves the way for a systematic approach to model 
heat transport in solids using multiscale modeling, in which the relaxation
time due to anharmonic 3-phonon processes is calculated quantitatively, in
addition to the usual harmonic properties such as phonon frequencies and 
group velocities.
It also allows the construction of accurate bulk interatomic potentials database.

\end{abstract}

\maketitle


\section{Introduction}
Classical molecular dynamics (MD) simulations use either semi-empirical potentials 
such as Stillinger-Weber (SW)\cite{sw}, Abell-Tersoff-Brenner\cite{atb}
or other type of force fields where the potential energy is an analytical function of 
the atomic positions, or first-principles potentials calculated typically using 
density-functional methods based on either the Born-Oppenheimer\cite{bo} or the Car-Parrinello\cite{cp} dynamics. 
The former are fast to compute but suffer from inaccuracies, while the latter are accurate but time-consuming to compute. 
Due to recent interest in thermal transport in semiconductor materials having good thermoelectric properties
and the topic of microelectronics thermal management in general, there have been many calculations of the lattice thermal 
conductivity of materials using the Green-Kubo (GK) formula\cite{green,kubo}. This formula relates the thermal conductivity,
through the use of the fluctuation-dissipation theorem, to the time-integral of the heat current autocorrelation
function. The latter is calculated from an MD simulation, and the ensemble average is usually 
replaced by a time average. Semi empirical potentials such as SW are usually used to perform the 
MD simulation for a system such as Si. As the thermal conductivity of a perfect crystal is mainly 
due to anharmonic three-phonon processes, directly related to the third-derivatives of the potential
energy with respect to atomic displacements, and the latter is generally not fitted or considered 
in the design of the semi-empirical potentials, there is really no good reason to expect an accurate
value for the thermal conductivity calculated from a  GK-MD simulation. In the case of Si, when using the 
SW potential, however, for some reason\cite{magic}, relatively good agreement is found between the 
experiment and the simulation results, even for a relatively small supercell\cite{volz,philpot,ase,sellan}.
The latter fact is also cause for concern, because, as we will show in the following, a small supercell
limits the number of long-wavelength phonons which carry a large portion of the heat in a material.
The lucky agreement can be attributed to a cancellation due to two different effects, which
will be discussed in section\ref{sizescaling}. 
Sellan {\it et al.}\cite{sellan} have presented a discussion on convergence issues, mostly with 
respect to non-equilibrium MD simulations, and we will also discuss the scaling issue 
with respect to GK-MD and lattice dynamics (LD) in this paper.

More accurate calculations of the thermal conductivity, based on the full solution of 
Boltzmann transport equation have shown that the thermal conductivity of Si using 
the SW potential is about 4-times larger than the 
experiment, while using the Tersoff or EIDP potentials produce results that are 
about twice larger than the experimental values\cite{broido05}.
In a similar work, using the environment-dependent interaction potential (EDIP), 
Pascual-Gutierrez {\it et al.}\cite{murthy1} also find 
thermal conductivity of bulk Si from MD in good agreement with experiments. In a subsequent
work\cite{murthy2}, the same group computed the thermal conductivity using the 
lattice dynamics (LD) theory based on the same EDIP potential, similar to 
the work of Broido {\it et al.} \cite{broido05}.
They obtain good agreement with experiments whereas Broido  {\it et al.} 
do not. The reason for this discrepancy, as they also mention in their paper, is unclear. Opinions on the accuracy of semiempirical potentials seem to differ as
there have been other reports\cite{70} where SW is found to overestimate the 
thermal conductivity by 70\%.
As we will show in this paper, some of these potentials might not be completely reliable 
for the calculation of the thermal transport properties for the simple reason that
they were not fitted or constructed to have the correct third derivatives, which 
are responsible for the thermal resistivity of a material. In fact even their 
harmonic force constants produce phonon dispersions and elastic constants which differ 
from experiments by 10 up to 40\%. Furthermore such potentials exist and 
have been thouroughly tested for only a very small number of pure crystalline solids.

In a tour de force work, Broido {\it et al.}, later, used the density-functional perturbation theory (DFPT) formalism in order to 
calculate the phonon scattering rates from first-principles DFT calculations and were able to 
successfully reproduce the thermal conductivity of bulk Si and Ge\cite{broido08}.
Their approach, which was very accurate, included  the calculation of all the cubic force
constants up to 4 lattice parameters away, and the complete iterative solution of the Boltzmann
transport equation.

We recently developed a methodology to extract second, third and fourth derivatives of the 
potential energy from first-principles calculations\cite{keivan-stokes},
and showed that the phonon dispersion relation in Si can be well-reproduced. In this paper, we pursue this 
work further and use these derivatives to construct a force field in order to explore the results from MD 
simulations and perturbation calculation to calculate the thermal conductivity of bulk Si. 
We should mention that our approach, even though very similar in essence to that of Broido {\it et al.},
is simpler in the sense that it limits  the range of the force constants (FCs) to a few neighbors
(5 for harmonic and 1 for cubic in the present case study of Si, in contrast to more than
20 for harmonic and 10 for cubic in the work of Broido {\it et al.}\cite{broido08}). For the 
sake of physical correctness, however, we enforce the translational, rotational and Huang invariances
on the extracted force constants. So the latter are not exactly equal to the ones obtained 
from DFPT or any finite difference calculation of the forces, but make the calculation load much lighter
than if one had to include so many neighbors. On the other hand a DFPT calculation, if restricted
to few neighbors, should enforce all these invariances. Usually, only the translational ones, also
known as the ``Acoustic Sum Rule" (ASR) are enforced in some of the standard DFT codes such as 
Quantum Espresso\cite{qe}.

In what follows we briefly review the methodology to extract force constants from 
first-principles density-functional theory calculations (FP-DFT).
The formalism for the molecular dynamics and Green-Kubo calculations of $\kappa$ are 
explained in section (\ref{mdgk}). This will be followed by the lattice dynamics approach
detailed in section (\ref{ld}). Results for Si will be shown and discussed in section (\ref{results}),
followed by conclusions.

\section{Extraction of force constants from FP-DFT calculations}

We construct the potential energy $V$ for the MD simulation as a Taylor expansion up to fourth order
in the atomic displacement $u_i$ of atom $i$ about its equilibrium position:

\beqnar
V&=&V_0 + \sum_i \Pi_i u_i +  \frac{1}{2!} \sum_{ij}\Phi_{ij} \, u_i u_j \\ & +&  
\frac{1}{3!} \sum_{ijk}\Psi_{ijk} \, u_i u_j u_k  +
\frac{1}{4!} \sum_{ijkl}\chi_{ijkl} \, u_i u_j u_k u_l \nonumber \\
&=& \sum_i (e_i - {1 \over 2} m_i v_i^2)
\label{potential}
\eeqnar

The last equation defines the on-site energy $e_i$ . If the displacements $u_i$ are 
around the equilibrium position, and this is usually the case, $\Pi_i=0$.
The coefficients $\Phi, \Psi$ and $\chi$ in the expansion are called the harmonic, cubic, and
quartic force constants respectively, and satisfy certain symmetry constraints. Namely they 
must be invariant under interchange of the indices, uniform translations and rotations of the atoms,
in addition to  invariance under symmetry operations of the crystal. 
The details of the needed constraints and how they are 
imposed can be found in our previous work\cite{keivan-stokes}. 

To get these numbers, we consider one or several supercells in which atoms are in their equilibrium
position. One, two or three neighboring atoms are moved simultaneously by a small amount,
typically about $0.01 \AA $ along the cartesian directions. Consideration of crystal symmetry 
usually reduces the needed displacements. For instance in a cubic-based crystal, like silicon,
where the two atoms in the primitive cell are equivalent, one only needs to move one Si atom
along the x direction. This is sufficient to extract all harmonic force constants if the supercell
size is large enough. The latter size is chosen depending on the available computational power and 
the considered range of force constants. 
To get three- and four-body interaction terms, one needs to move two and three atoms at a time and 
record the forces on all atoms in the supercell. 
It is advantageous to record forces for atomic displacements in two opposite directions as there
would be cancellation of the cubic contributions and this would make the calculation of the
harmonic FCs much more accurate (up to order $u^2$).
Thus one would obtain a large set of force-displacement relations computed from a FP-DFT code.
Together with the invariance constraints, an overcomplete linear set of equations on all
the force constants will be formed. A singular-value decomposition algorithm is then used to solve this 
linear overcomplete set. We find that usually the violation of the invariance relations is 
of the order of $10^{-6} $ times the FC itself. This however requires very accurate evaluation
of the forces, meaning that they should have converged with respect to the cutoff energy and 
number of k-points to within at least 4 significant figures, if not more!
Our experience on graphene\cite{mingo} and silicon (present work and reference [\cite{keivan-stokes}] 
has shown that the harmonic force constants are usually reproduced quite accurately. Higher order
FCs have less accuracy as their contribution shows up in the second or third or fourth significant figures
of the forces. The main approximation is in cutting off the range of the interactions, which
will lead to inaccuracies in some of the Gruneisen parameters as we will shortly see.

\section{Molecular dynamics and the Green-Kubo formalism}
\label{mdgk}

Typically a supercell is constructed with periodic boundary conditions, 
and an MD simulation is performed over a long enough time steps in order to 
reach thermal equilibrium, followed by a long (N,V,E) simulation in order
to collect data on $J$ for later statistical processing, i.e time and ensemble averaging
of its autocorrelation. 

Based on the potential displayed in Eq. (\ref{potential}), one can extract the expression
for the force, required in the MD simulations:

\beq
F_i 
= - \sum_j  u_j \, (\Phi_{ij}  + \frac{1}{2!} \sum_{k}\Psi_{ijk} \, u_k  +
\frac{1}{3!} \sum_{kl}\chi_{ijkl} \, u_k u_l )  \nonumber
\label{force}
\eeq

The heat current is defined in the discrete (atomic) case of a lattice, where there is no convection, as:

\beq
J^{\al} = \sum_i J_i^{\al} = \sum_i {d (e_i r_i^{\al}) \over dt} = \sum_{ij} (R_i^{\al}-R_j^{\al}) (v_j \cdot \, {\partial e_i \over \partial u_j } )
\label{gen-current}
\eeq
where $v_i$ is the velocity of particle $i$ and $e_i$, as defined in Eq. (\ref{potential}) is the 
local energy of atom $i$. Using our expansion, it can be expressed as a function of the 
force constants as follows (as we expand around the equilibrium position, we assume $\Pi_i=0$) :

\beqnar
e_i &=& {1 \over 2} m_i v_i^2 + {u_i \over 2} \sum_j  [ 
 \Phi_{ij} \, u_j +  \frac{1}{3} \sum_{k}\Psi_{ijk} \, 
 u_j u_k  \nonumber \\
 &+& \frac{1}{12} \sum_{ijkl}\chi_{ijkl} \, u_j u_k u_l  ] 
\label{local-energy}
\eeqnar

This definition leads to the following form of the local heat current:

\beqnar
J_i^{\al} &=& {1 \over 2} \sum_{j} (R_i^{\al}-R_j^{\al}) (v_j \cdot u_i [ 
\Phi_{ij} +  \frac{2}{3} \sum_{k}\Psi_{ijk} \, u_k \nonumber \\
&+& \frac{1}{4 } \sum_{kl}\chi_{ijkl} \, u_k u_l ] 
\label{current}
\eeqnar

Finally the thermal conductivity tensor is given by the well-known Green-Kubo 
relation\cite{green,kubo}:

\beq
\kappa_{\alpha \beta} = {1 \over V k_B T^2} \int_0^{\infty} \, < J^{\alpha}(0) J^{\beta}(t) > \, dt
\eeq
where $\alpha,\beta=x,y,z$, and $< A >$ denotes the equilibrium average of the observable $A$, 
which in the classical case
can be replaced by its time average provided the time is {\bf long enough to satisfy ergodicity}.
The ensemble averaging is necessary as long as different runs starting with different initial
conditions lead to different integrated autocorrelation functions.
The true current autocorrelations decay quite fast. 
In a single MD run, however, this decay is not observed. Instead one observes a decay in the 
amplitude followed by random oscillations about zero.
As finite systems are usually not ergodic, an ensemble average, over the random initial conditions
is also needed to correctly simulate the equilibrium average required in the Green-Kubo formula.
In this case, the ensemble-averaged autocorrelation function will decay smoothly to zero with time. 
One will then see that the decay is indeed relatively short,
because the long-time tails get cancelled after ensemble averaging.
The advantage of the ensemble averaging, in addition of course to the usual time averaging, is that one
samples the phase space more randomly, and generates {\bf uncorrelated} sets of pairs 
$J^{\alpha}(0) J^{\beta}(t)$ for a given time difference $t$. 
In this case, the mean has the convergence properties of gaussian-distributed variables 
and the error decays to zero as the inverse square root of the number of initial conditions.

In a numerical simulation, the GK formula should be replaced by:

\beqnar
\kappa_{\alpha \beta} &=& {1 \over V_{\rm cell} \, k_B T^2} {1 \over N_{\rm ens}} 
\sum_{i=1}^{N_{\rm ens}} \, 
{\Delta t \over \T}  \nonumber \\
& [ & \sum_{t=0}^{\T} \sum_{p=1}^{\T-t+1}   \,  J^{\alpha}_i(p) J^{\beta}_i (p+t) ]
\label{GK-numerical}
\eeqnar
where $\T$ is the total simulation time of each run, $\Delta t$ is the time difference between two successive data points,
$t$ and $p$ are integers labeling time,
and $N_{\rm ens}$ is the number of generated initial conditions, each labeled by $i$.
One important comment is in order here, and that is the use of $1/\T$ in the denominator instead of 
the  more intuitive $1/(\T-t+1)$ which is the actual number of terms in the last sum.
One can show that the choice in Eq. (\ref{GK-numerical}) provides an unbiased estimator of the autocorrelation\cite{statbook}.
Some previous works in the GK method such as reference \cite{philpot} have used the ``biased" formula, 
which would be 
fine as long as the total simulation time is much larger than the largest time $t$ used for the integration.
One simple way to become convinced that Eq. (\ref{GK-numerical}) is the correct way, is to consider large times $t$ near the 
maximum simulation time $\T$. 
As $J(p) $ and $J(p+t)$ are both fluctuating random numbers, and there are not 
enough terms in the sum to make the average go to zero, the time average will 
become large again and tend to $\sigma_J=<J(0) J(0)> $ instead 
of decaying to zero as $t$ approaches $\T$. A division by the small number $\T-t$ would not solve the problem, 
whereas the division by $\T$ would make this very small as it should.

The simulation time $\T$ being usually quite long, the statistical error in the time averging is
usually small, and we estimate the error bars in our data from the ensemble averaging:
If ${\bar C}(t)$ is the ensemble-averaged autocorrelation function, its error bar $\Delta C$ is evaluated as:
$\Delta C(t)= [\sum_{i=1}^{N_{\rm ens}} (C_i(t)-{\bar C}(t))^2]^{1/2}  /N_{\rm ens}$

The magnitude of this error bar and the required accuracy in the results determine how
many ensembles are needed for a proper estimation of thermal conductivity. 

\subsection{How many MD steps are necessary?}

For a given supercell size, there is a discrete number of phonon modes which can propagate and
get scattered in the system. The largest wavelength consistent with the periodic boundary
conditions would be the supercell length. To
this, one can correspond a smallest phonon wavenumber or frequency allowed in the simulation: $\omega_{cut}=2 \pi c/L$.
The total simulation time should be large enough so that all phonon modes can get scattered 
a few times within the simulation period.
Largest relaxation times belong to acoustic modes, and usually decay as the inverse
square of the phonon frequency. Knowing the smallest allowed frequency $\omega_{cut}$ due to the finite
size of the system, one can estimate the corresponding phonon lifetime (using Klemens'
formula displayed in Eq. (\ref{klemens}) for instance). 
The total MD simulation time should therefore be a few times larger
than the largest phonon lifetime so that scattering events of long wavelength phonon modes 
can properly be sampled during the MD run. As an example, we can consider Si system
in a cubic 10x10x10 supercell of 8000 atoms. In this case $L=10 \times 5.4 \AA=5.4$ nm
corresponding to $k_c= 2 \pi/10 a$ which is a fifth of the $\Gamma \to X$ line.
The lowest frequency mode is therefore about $\om_{cut}= 1 $ THz.
As can be seen in Fig. (\ref{gama}) below, the  
normal and umklapp lifetimes at this frequency are 3000 ps and 10000 ps respectively,
leading to a total lifetime of 2300 ps. So in order to sample such rare events, one
needs to calculate the autocorrelation over at least 20 ns, which means the 
MD simulation should be run for at least the same amount of time if not longer.
This could be computationally prohibitive. If the runs are made with fewer MD steps
long wavelength phonons will not be relaxed and the autocorrelation would not tend to zero.
Another consequence of this remark is that for large supercells, as 
relaxation times scale as $1/\om_{cut}^2 \propto L^2$,  
longer simulation times proportional to the {\bf square} of the 
the supercell size would be required.

\subsection{Size scaling}
\label{sizescaling}

As mentioned, the choice of a finite supercell comes with the cost of discretizing the 
phonon modes and supressing 
the phonons of wavelengths longer than the supercell length. The neglected contribution 
maybe estimated as follows: the anharmonic 
lifetime of acoustic modes maybe approximated by the Klemens' formula \cite{klemens}
\beq
{1 \over \tau_{\kl}^{Klemens}} = \gamma_{\kl}^2 \, {2 k_B T \over M \, v_{\kl}^2} \, 
{\omega_{\kl}^2 \over \omega_{\lambda}^{max}}
\label{klemens}
\eeq
where $\omega_{\lambda}^{max}$ is the largest frequency of the branch $\la$,
$\gamma_{\kl}$  the mode Gruneisen  parameter, $\omega_{\kl}$ the 
frequency, and $v_{\kl}$ the group velocity associated with the mode $\kl$.
Therefore long wavelength phonons will have a large relaxation time and can
considerably contribute to the thermal conductivity.
Assuming this form in the relaxation time approximation to the thermal 
conductivity, and using Eq. (\ref{rta}), we can write the thermal conductivity
as a sum over contributions of phonons of different frequencies:
$$
\kappa = \int_0^{\infty} {1 \over 3} \tau(\om) v^2(\om) C_v(\om) DOS(\om) \, d\om
$$

In 3D, since the density of states (DOS) is quadratic in frequency, 
the contribution
of long wavelength acoustic phonons would be linear in the cutoff frequency $\om_{cut} = 2\pi c/L$:
$$
\kappa(L) = \kappa(\infty) - \int_0^{\om_{cut}} {1 \over 3} \tau(\om) v^2(\om) C_v(\om) DOS(\om) \, d\om
$$
For low frequencies $C_v(\om)=k_B [\beta \hbar \om/2 {\rm sinh} (\beta \hbar \om/2) ]^2 
\simeq k_B$ and $v(\om) \simeq c$ so that
\beqnar
\kappa(L) &=& \kappa(\infty) - A \int_0^{\om_{cut}} {1\over \om^2}  DOS(\om) \, d\om = \kappa(\infty) - 
D \om_{cut} \nonumber \\
&=& \kappa(\infty) - E {1 \over L} = \kappa(\infty) - F/\sqrt{\Lambda_c}  
\label{scaling}
\eeqnar
where $A, D, E$ and $F$ are constants which do not depend on the size, $\Lambda_c$ is the 
mean free path (MFP) associated with the cutoff frequency  $\om_{cut} = 2\pi c/L$.
This gives us a way to deduce how the  thermal conductivity of a finite size sample 
scales with the supercell length or the cutoff frequency $\om_{cut}$ or MFP, $\Lambda_c$.
We should note that a different scaling law ($1/\kappa(L) =1/\kappa(\infty) + C / L$) was also
proposed and used by Sellan {\it et al.}\cite{sellan}, and also Turney {\it et al.}\cite{turney}
when they want to extrapolate their thermal conductivity data to infinite size. 
The argument they used to deduce it 
however was based on the Matthiesen's rule, stating that the bulk resistivity is obtained 
by adding the finite size resistivity to the one obtained from $L/v$ taken as relaxation time. 

There is another additional problem with finite size MD simulations. Even though momentum
is still conserved in a 3-phonon process, because the modes are {\bf discrete} in a 
finite supercell, energy
conservation will not always be possible, unless the energy difference 
$\om-\om_1-\om_2 \le \Gamma$
where $\Gamma$ is on the order of the sum of inverse  lifetimes of the three considered phonons.
If this relation is not satisfied, the considered 3-phonon scattering will not take 
place in a finite supercell, and this will lead to an overestimation of the lifetime 
of the phonons, and thus, of the thermal conductivity. 

These competing effects, namely an overestimation of $\kappa$ due to limited phase space for energy conservation 
and an underestimation due to cutoff of low frequency acoustic modes, may lead to a 
magical cancellation, resulting in thermal conductivities in good agreement with
experiments even for moderate supercell size. This error cancellation will likely
affect the temperature-dependence of $\kappa$: at higher temperatures the discreteness
error is reduced as $\Gamma$ increases linearly with $T$. The frequency cutoff
error, however, will not be affected by high temperatures. Consequently, as $T$ is increased
the thermal conductivity of a finite sample will decrease faster than $1/T$ with temperature. This has been observed in the work of Volz and Chen\cite{volz}.
It can also be verified by introducing other scattering events such as isotope or 
defect scattering
leading to larger $\Gamma$ values. In such cases, the discreteness of modes will have  
little effect, and the simulated $\kappa $ will be 
less than the exact one, due to the cutoff of long MFP phonons effect. As a 
result, in a system where due to disorder or high temperatures scattering 
rates are high, GK-MD simulations
will typically require larger supercells to converge.

The correct way of estimating $\kappa(T)$ is to do a proper size scaling at each temperature
by plotting $\kappa(T,L)$ versus $1/L$ and linearly extrapolating to $1/L \to 0$. 

\section{The Lattice Dynamics approach}
\label{ld}

Using the extracted for constants, one can form the dynamical matrix of the crystal
using its primitive cell data:
\beq
D_{\tau \tau'}^{\alpha \beta} (\k) = \sum_{\R} {1 \over \sqrt{M_{\tau} M_{\tau'}}} \,
\Phi_{0 \tau,\R \tau'}^{\alpha \beta} \, e^{ i \k \cdot \R}
\eeq
where $\R$ is a translation vector of the crystal, $\tau$ refers to an atom in the
primitive cell, and $\alpha,\beta$ are cartesian components $x,y,z$.
Such sums of the force constants over the translation vectors of the primitive lattice are
usually short-ranged and fast to compute, except if Coulomb interactions are involved, in which
case the sum is evaluated using the Ewald method.

Diagonalizing this matrix, one can find the phonon spectrum and the normal modes as its eigenvectors:
\beq
\sum_{\tau' \beta} D_{\tau \tau'}^{\alpha \beta} (\k) \, e_{\la}^{\tau' \beta} (\k) = 
\omega^2_{\kl} \, e_{\la}^{\tau \alpha } (\k)
\eeq
where $\la$ labels a phonon band (or branch), and $k$ refers to a point in the first Brillouin zone (FBZ).
Using these eigenvectors and eigenvalues, and from perturbation theory, one can 
calculate the phonon
lineshifts and lifetimes as the real and imaginary parts of the 3-phonon self-energy defined as\cite{marad,cowley,srivastava,reissland}:
\beqnar
\Sigma(\ql,\omega) = - {1 \over 2 N_k} \sum_{1,2,\ep=\pm 1} |V(\ql,1,2)|^2 \, \times \nonumber \\
\left[ { (1+n_1+n_2)  \over
\om_1 +\om_2 + \ep \omc} 
+ { (n_2-n_1) \over \om_1 -\om_2 + \ep \omc} \right]
\label{self}
\eeqnar
where $\omc = \om-i \eta $ , $(\eta \simeq 0^+) $ is a small infinitesimal number, 
which in practice is taken to be finite for a given k-mesh size, n is the equilibrium 
Bose-Einstein distribution function, and 1 and 2 refer to modes $(\qo )$ and $(\qt )$.
The 3-phonon matrix element $V$, expressed as a function of the cubic force constants $\Psi$, 
is given by:
$$ 
V(\ql,1,2) = ({ \hbar \over 2})^{3/2} \sum_{ \R_i  \tau_i \al_i }  
\Psi^{\alpha \beta \gamma}_{0 \tau, \rto , \rtt } \times $$ 
\beq 
\frac{ e^{i (q_1 \cdot \R_1 + q_2 \cdot \R_2)} \, e_{\la}^{\tau \alpha } (q) \, e_{\la_1}^{\tau_1 \alpha_1 } (q_1) \, e_{\la_2}^{\tau_2 \alpha_2 } (q_2)} {
[M_{\tau} M_{\tau_1} M_{\tau_2} \, \om_{\ql} \, \om_1 \, \om_2 \, ]^{1/2} }
\eeq
The calculation of the self-energy would require a double sum over the 
q-points (labeled above by 1 and 2) in the FBZ . 
Due to the conservation of momentum, however, only terms with $q+q_1+q_2=G$, 
with $G$ being a reciprocal lattice vector, should be included in the above sum. 
In practice, therefore, this involves only a single summation. 
To get the phonon dispersion and lifetimes due to 3-phonon scattering terms, one needs
to solve $E = \om_{\kl} + \Sigma'(\kl,E)$ where $\Sigma'$ 
is the real part of the self-energy ($\Sigma=\Sigma'+i\Sigma"$) . 
This equation needs to be solved iteratively. Since the shift is usually small, to leading
order, one can use $E = \om_{\kl} + \Sigma'(\kl,\om_{\kl})$ i.e. one uses the on-shell 
frequency as argument of the self-energy. The same approximation will be used for the imaginary
part giving the inverse lifetimes. 
The corresponding phonon lifetime will be given by $\tau_{\kl} = 1/ 2 \Sigma" (\kl,\om_{\kl}) $.
In the evaluation of the imaginary part $\Sigma" $, one encounters  Dirac delta functions
reflecting the conservation of energy in the three-phonon process: $\om_{\kl} = \om_1 \pm \om_2$.
In effect, from Eq. (\ref{self}), it can be noticed that the delta function is substituted 
by a Lorentzian 
function of width $\eta$. The latter depends on the choice of the k-point mesh in the FBZ. A small value for 
$\eta$ can be used for a fine mesh, while a coarse mesh requires larger values of $\eta$. Typically
$\eta$ is chosen to be of the order of energy spacing in the joint density of states (JDOS) so 
that the latter is a smooth function of the frequency and does not display any oscillations with
sharp peaks which would appear if the width is too small.
\beq
JDOS(\om) = {1 \over N_k}  \sum_{1,2} \delta(\om - \om_1 -\om_2) + \delta(\om - \om_1 + \om_2)
\eeq

The anharmonicity can be characterized by the Gruneisen parameters (GP). 
The force constant GP is defined as $\gamma_{\phi} =$ -d ln $\phi/2$ d ln $V$ where $V$ 
is the volume. The mode GP is defined as:
$\gamma_{\kl}$ = -d ln $\omega_{\kl}/$ d ln $V$ where $\omega_{\kl}$ is the phonon 
frequency evaluated at the point ${\vec k}$ and band index $\la$.
It gives the relative decrease in the phonon frequency as the volume is increased by 1\%.
From the Taylor expansion of the harmonic force constants in terms of the volume 
or the lattice parameter, one can calculate such change.

\beqnar
\gamma_{\kl} &=& -{1 \over 6 \om_{\kl}^2 } \sum_{1,2} 
\Psi^{\alpha \alpha_1 \alpha_2}_{0\tau,\rto,\rtt} \, 
{e^{i k \cdot (\R_2 - \R_1)} \over
 [ M_{\tau_1} M_{\tau_2}  ]^{1/2} } \nonumber \\
 &\times & \, X_{0 \tau}^{\alpha }  \, e_{\la}^{\tau_1 \alpha_1 } (-k) \, e_{\la}^{\tau_2 \alpha_2 } (k)
\label{gruneisen}
\eeqnar
where $X_{R \tau}$ is the equilibrium atomic position of atom type $\tau$ in the primitive cell labeled by
the translation vector $\R$.

Finally, the thermal conductivity is calculated within the relaxation time approximation (RTA),
which leads to the following well-known expression for the thermal conductivity:
 
\beq
\kappa = {1 \over 3 \Om N_k} \sum_{\kl} v_{\kl}^2 \tau_{\kl} \, \hbar \omega_{\kl} \, \partial  n_{\kl} /\partial T 
\label{rta}
\eeq
where $\Om$ is the volume of the unit cell.
The relaxation time $\tau_{\kl}$ in this expression represents the time after which 
a phonon in  mode $\kl$ reaches equilibrium on the average, and depends on the 
scattering processes involved. In a pure bulk sample, the only source of phonon
scattering is anharmonicity dominated usually by three-phonon processes. 
Using perturbation theory or the well-known Fermi Golden rule (FGR), one can derive the 
expression of the relaxation time as a function of the cubic force constants\cite{marad,cowley,srivastava,reissland}. 
It can be shown that to a good approximation, it is given by 
\beq
\tau_{\ql} \approx {1 \over  2 \Im [\Sigma(\ql,\om_{\ql})] }
\eeq
In what follows, we have disregarded the boundary scattering term, which is responsible for the low-temperature behavior of $\kappa$. In such case, $\kappa$ is expected to saturate to a finite value at low enough temperatures. The reason for this saturation can be understood if one assumes
the low-frequency limit of the DOS and the relaxation times similar to Eq. \ref{scaling}. 
Considering that in $\om \to 0$ limit we have: 
$DOS_{\la}(\om) \to \om^2/2 \pi^2 c_{\la}^3 $ 
and $C_v(\om) = k_B (x / {\rm sh} x)^2$ (with $x=\beta \hbar \om/2 $), and
the relaxation time can be written as $\tau(\om) \to \hbar \om_o^2/\om^2 k_B T$, 
the integral defining the thermal conductivity can be transformed, in the low
temperature limit, to:
$$
\kappa(T) = \frac{ k_B \om_o^2}{\pi^2 c_{\la}}   \int_0^{\infty} (\frac{x}{{\rm sh} x})^2 \, dx = \frac{ k_B \om_o^2}{6 c_{\la}} 
$$ 
The constant $\om_o$ that appears in the low energy limit of the relaxation 
time as well as the speeds of sound $c_{\la}$ determine the saturated value of the thermal conductivity. 
So when only 3-phonon scattering processes are included, the thermal conductivity would
tend to $ k_B \om_o^2/6 c_{\la}$  as $T$ goes to 0.

Finally, in our numerical calculations where the integral in the 
FBZ has been approximated by a sum over a discrete set of k-points,
the low-frequency region is not properly sampled and
we observe a decay to zero at low $T$, and therefore have not reported
the unreliable low-temperature data in this work.

\section{Results and Discussions}
\label{results}

\subsection{Validation of force constants}

First-principles calculations were done using the PWSCF code of the Quantum Espresso package\cite{qe}
A set of force-displacement data were calculated using 
$2\times2\times2$ supercell of 64 Si atoms. The set of force-displacements data, along 
with the symmetry constraints, form an over-complete linear set of equations needed to determine the 
potential derivatives. We use the local density approximation (LDA)
of Perdew and Zunger\cite{pz} with a cutoff energy of 40 Ryd and 10 k-points in the irreducible
Brillouin zone of the cubic supercell. The range of different ranks of force constants can be chosen
by the user. We have set the range of harmonic forces constants (FCs) to 5 nearest neighbor shells, 
and that of the cubic and quartic force constants to the first neighbor shell only.
This results in 17, 5 and 14 independent harmonic, cubic and quartic FCs respectively.
The corresponding number of terms in the Taylor expansion of the potential energy are, however,
equal to 1500, 1146 and 7980 respectively. This is why the ranges were restricted to 5, 1 and 1 nearest
neighbor shells in order to limit the computational time to a reasonable amount. Note that despite
the large number of terms to be computed, arithmetic operations are only limited to additions and 
multiplications.

\begin{figure}[h]
\includegraphics[angle=270,width=0.45\textwidth]{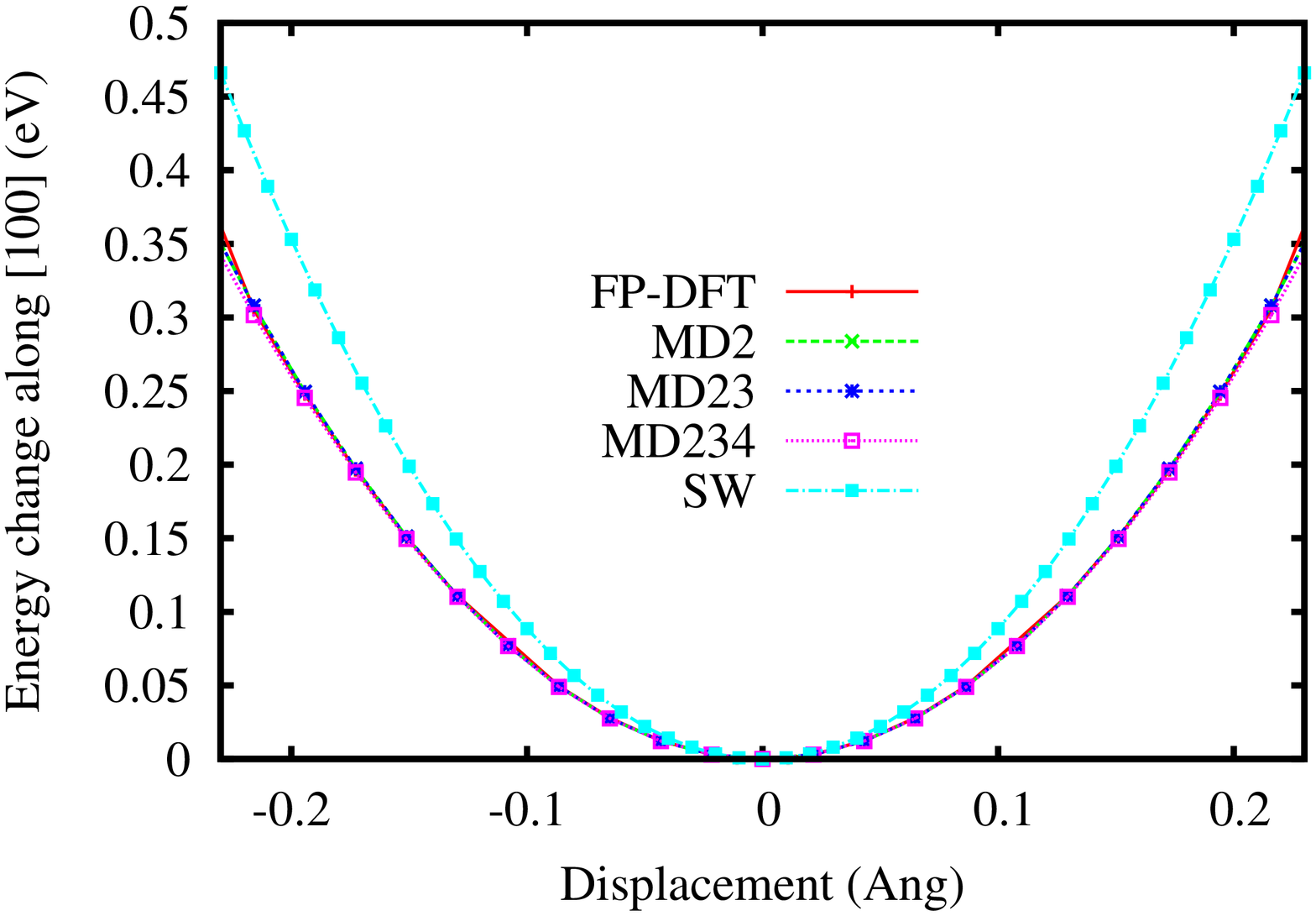}
\includegraphics[angle=270,width=0.45\textwidth]{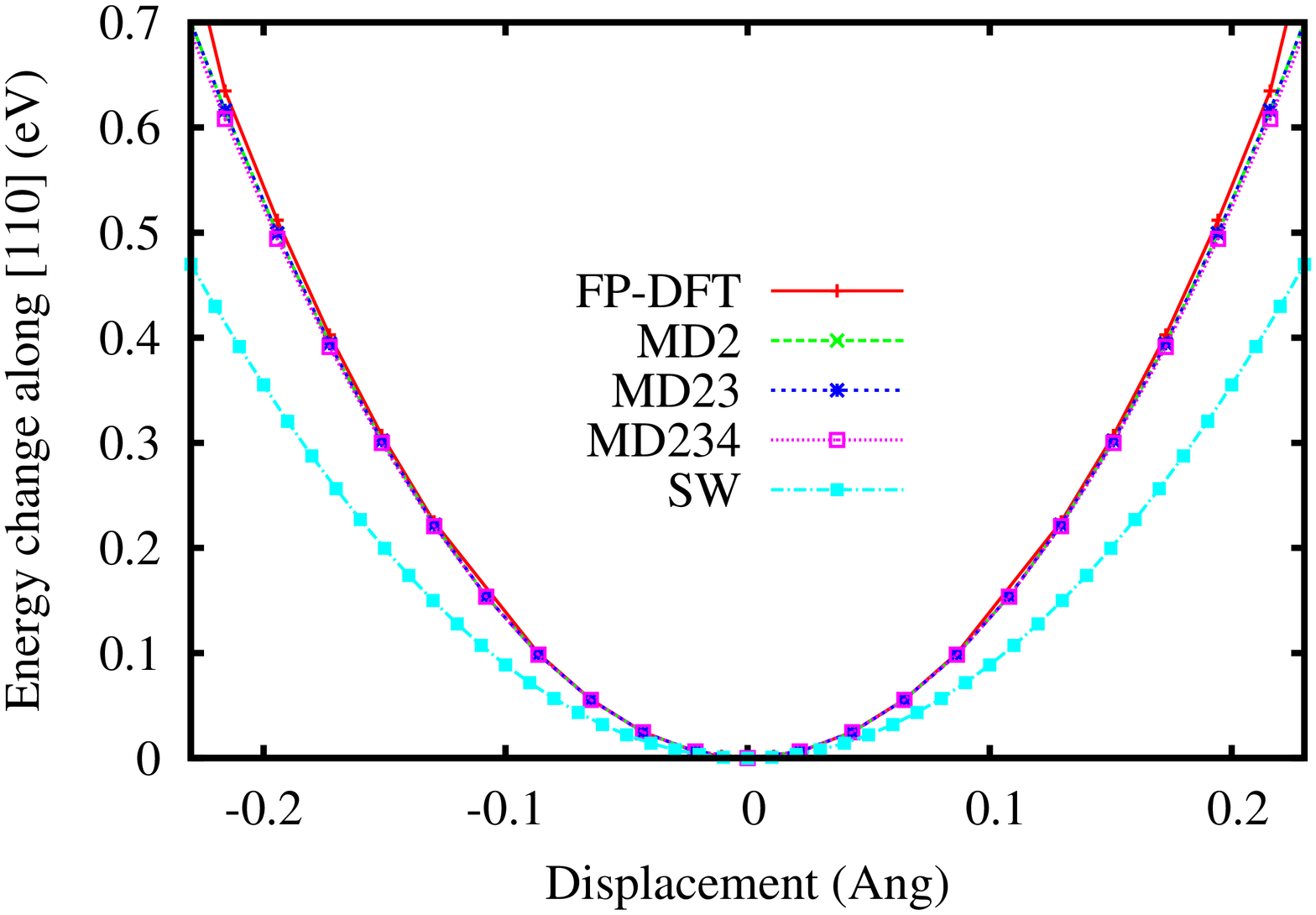}
\includegraphics[angle=270,width=0.45\textwidth]{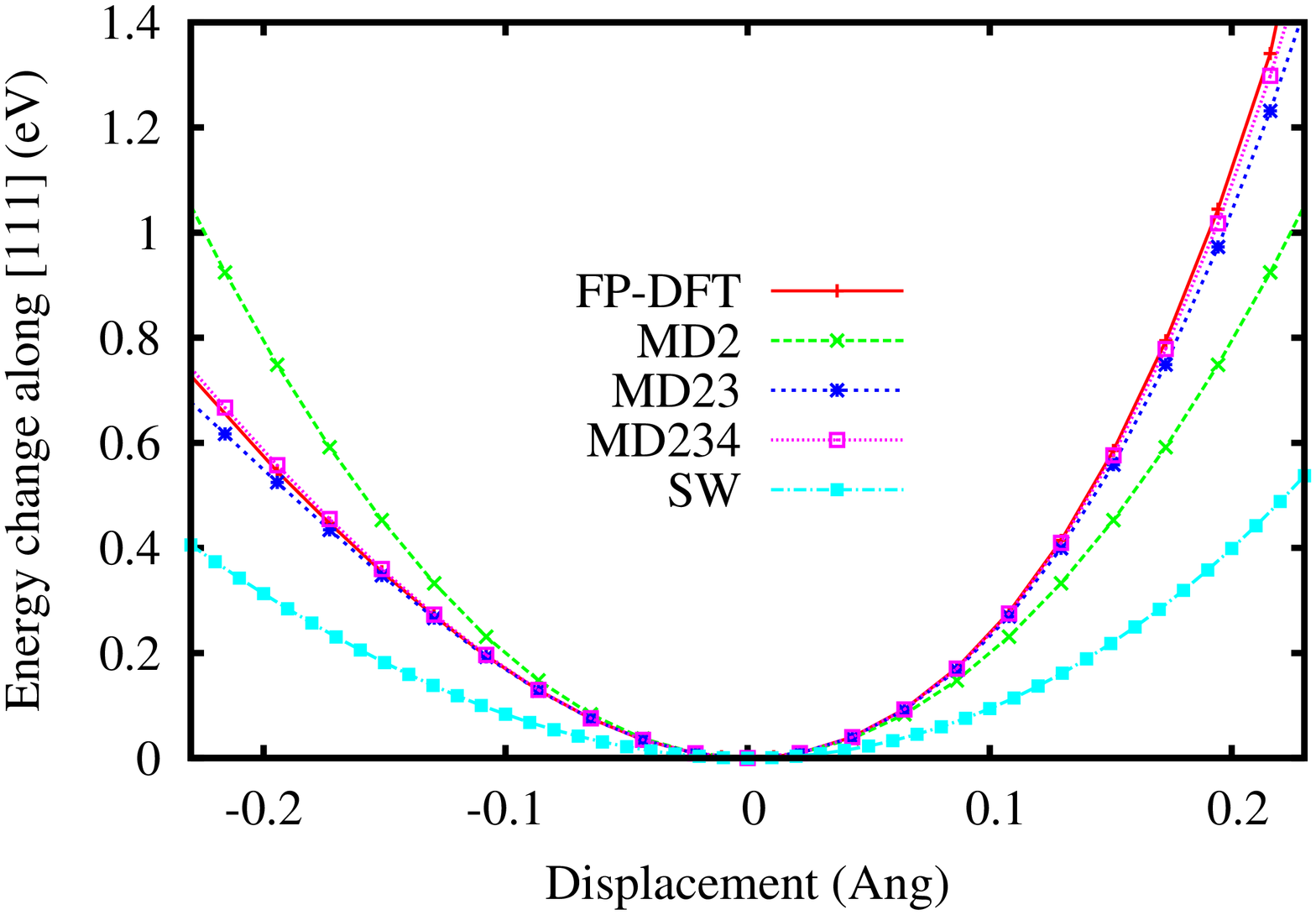}
\caption{(Color online) Total energy as an atom is moved in the [100] (left), [110] (middle) and [111] (right) 
directions. DFT results are compared with the force field and the Stillinger-Weber potential. MD234 refers to the force field in which all harmonic, cubic and quartic terms are included, while MD2 refers only to the harmonic force field, etc. 
}
\label{directions} 
\end{figure}

In Fig. (\ref{directions}), we show the change in the total energy as an atom in the supercell
is moved along the [100], [110] and [111] directions  respectively. Resutls from DFT calculations
are compared against our developed force field including the harmonic, harmonic+cubic, and
harmonic+cubic+quartic terms of the Taylor expansion. For the sake of comparison, we have 
also plotted the same energy change ontained from the Stillinger-Weber (SW) potential\cite{sw}, which
is widely used in MD simulations of Si systems.

To further assess the accuracy of the force field, we have also moved all the atoms
in the supercell in different random directions by a small amount of magnitudes 0.1 and
0.2 $\AA$ respectively, and compared the average force of our model and the 
SW potential to the FP-DFT one. The deviation is charaterized by:
\beq
\sigma({\rm model}) = {\sum_{i \al}  {( F_{i \al}^{\rm model} - F_{i \al}^{\rm DFT} )}^2 \over  \sum_{i \al} {F_{i \al}^{\rm DFT}}^2 }
\eeq
The results for the parameter $\sigma$ are summarized in table (\ref{errors}).

\begin{table}
\caption{Typical deviations in the SW and Taylor expansion (present model) force fields
compared to true FP-DFT forces. They are obtained by moving all 64 atoms in the
supercell in a random direction by 0.1 and 0.2 $\AA$ respectively.}
\begin{tabular}{|c|c|c|}
\hline
Amplitude($\AA$) &  $\sigma$(SW)  & $\sigma$(Present)   \\ 
\hline \hline
0.1 & 0.35 &  0.05  \\
0.2 & 0.28 &  0.08  \\
\hline
\end{tabular}
\label{errors}
\end{table}

We can notice that this type of error estimate would also include contributions from many-body forces, and is 
a more stringent test on the force field. The errors from the present model are consistently
about 4 to 5 times smaller that the SW potential.

In the following we follow two paths to compute the thermal conductivity.
The first is to use the Green-Kubo formula, by using the results from an MD simulation:

\subsection{Thermal conductivity from MD-GK}

\begin{figure}[h]
\includegraphics[angle=270,width=0.50\textwidth]{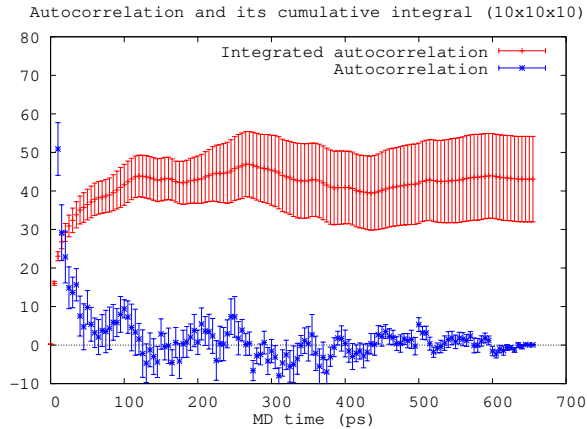}
\caption{(Color online) Plot of the ensemble-averaged (over 27 initial conditions) heat current 
autocorrelation as a function of time, and its 
integral for the 10X10X10 supercell. Vertical units for the integrated 
autocorrelation are in W/mK, and the autocorrelation (blue dots) has been 
multiplied by a constant to be on scale.
}
\label{autocor} 
\end{figure}

As previously mentioned, there will be large fluctuations in the current autocorrelation
function versus time from one run to the next, and therefore an averaging over several initial
conditions is necessary to produce a reliable plot.
In Fig. (\ref{autocor}), we have plotted such ensemble average for a 10x10x10
supercell containing 8000 atoms. The error bars are mainly due to the ensemble averaging,
and those related to the time averaging are small as the number of MD time steps are
quite large. 

We can also see in this figure the cumulative integral of the ensemble-averaged autocorrelation 
function. 
The same calculation was performed in a 7x7x7 supercell of 2744 atoms, where
the averaging was over 99 runs with different initial conditions. Due to its larger size,
there are smaller fluctuations in the average current per atom in the 10x10x10 supercell, 
and we only used 27 initial conditions for this supercell. 
Since from each MD run one can really 
extract three autocorrelation functions $\kappa_{xx}$,  $\kappa_{yy}$ and  $\kappa_{zz}$, 
which are equal by cubic symmetry, we also averaged over the 3 directions.
In this sense, the above mentioned numbers should be multiplied by 3.

The error bars are determined by the large fluctuations
in the integrated autocorrelations divided by the square root of the 
number of ensembles. The error bar due to the time average is usually 
much smaller if MD simulations are run for a long enough time.

The results for two different supercell sizes are summarized in table (\ref{kappa7-10}) as compared
with the experimental data of Slack {\it et al.}\cite{slack-si}. 
One can notice an underestimation of the 
experimental data, which is reduced as the supercell size is increased. 
To get the correct value in the thermodynamic limit, one needs to extrapolate 
these results to infinite size.

\begin{table}[h]
\caption{Thermal conductivity at T=600 K, from GK-MD compared to lattice dynamics in the classical limit
$(n(\om) \to k_BT/\hbar \om)$ with an equivalent number of k-mesh, 
and experiment for two different supercell sizes.} 
\begin{tabular}{|c| l| l| r|}
\hline
Supercell size & MD-GK & LD & experiment \\
\hline \hline
7x7x7 & 37 $\pm$ 10 & 32.67 & 64 $\pm$ 3 \\
10x10x10 & 43 $\pm$ 12  & 47.2  & 64 $\pm$ 3 \\
\hline
\end{tabular}

\label{kappa7-10}
\end{table}

There are a few competing effects which can explain this discrepancy:
the most important one is size effect, which as was just explained, underestimates $\kappa$. 
Similarly, the larger value of the Gruneisen parameter for the acoustic modes
in our model will produce a smaller relaxation time (see the Klemens formula in Eq.(\ref{klemens})). 

The following effects will, however, lead to an overestimation of the thermal conductivity:
in the classical MD simulations, the number of modes is the high-temperature limit of the Bose-Einstein distribution,
$k_BT/\hbar \omega_{\kl}$ which is larger than the quantum distribution. 
This leads to a heat capacity
per mode of  $k_B$ and therefore an overestimate of the true heat capacity (see also Fig. (\ref{kappa-cl})).  
In a finite size cell, the allowed frequencies are quantized and energy conservation after
a 3-phonon process can never be exactly satisfied, this will lead to an effectively longer
lifetime for phonons, and thus also overestimate $\kappa$.  It is not easy to quantify
these errors except for those due to the phonon occupation numbers. 
It is therefore possible that there is a cancellation. In our case, since only two supercell sizes
were considered, we can not do a systematic size scaling study, but overall, due to these
cancellations the MD-GK results seem to be weakly dependent on size, in agreement
with previous MD simulations (see for example Table I in reference [\cite{sellan}]).

Here, we must point out some discrepancy between published results on Si
using the SW potential.
Using the MD-GK method, Philpot {\it et al.} and Volz {\it et al.}  \cite{philpot,volz} find
a thermal conductivity in reasonable agreement (to within 30\%) with 
experiments.
Broido {\it et al.}\cite{broido05}, on the other hand, have shown by solving Boltzmann equation 
beyond the RTA,
that $\kappa_{SW} \geq\  \approx 4 \kappa_{\rm experiment}$. Recently Sellan {\it et al.}\cite{sellan}
investigated size effects in GK-MD simulations, direct method, and also used lattice dynamics to 
compute the thermal conductivity of Si from the SW potential. 
They found that $\kappa_{LD}(T=500K)=132 W/mK$, which is 
only 70\% larger than the experimental value of 80 W/mK, in contrast to Broido {\it et al}'s 
\cite{broido05} prediction. 
Their direct method followed by scaling predicts $93 \pm 18 $ W/mK, and
their unscaled GK value for a 8x8x8 supercell is ($231 \pm 57 $ W/mK).

All these results point to the subtleties involved in extracting a reliable value
for the thermal conductivity of bulk materials, no matter what method is used.

To investigate this discrepancy, we used our approach to extract cubic 
force constants from the SW potential and used LD theory to compute the
corresponding thermal conductivity. Using the same k-point mesh, in order 
to avoid systematic errors, in comparison to FP-derived force constants, we
found that at 150K the thermal conductivity derived from SW is 80\% larger 
than the one derived from FP-DFT calculations.

\subsection{Phonons, DOS and Gruneisen parameters}

\begin{figure}[t]
\includegraphics[angle=270,width=0.50\textwidth]{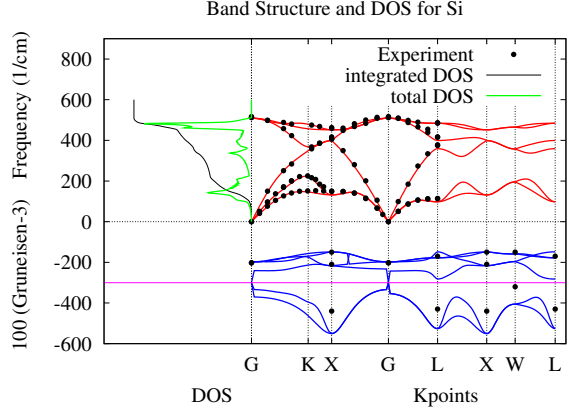}
\caption{(Color online) Phonon band structure of Si using force constants up to fifth neighbor shell. 
Plus signs represent experimental data of Nelin and Nilsson\cite{nilsson}. Left: DOS , and 
bottom: Rescaled Gruneisen parameters ($100 \times (\gamma-3)$) }
\label{dispersion} 
\end{figure}

In extracting the force constants, we have limited the range of the harmonic
FCs to 5 neighbor shells, and that of the cubic and quartic terms to one neighbor shell, so that
MD simulations can be done within a reasonable time. 
Using the harmonic FCs, we can obtain the phonon spectrum. 
As can be seen in Fig. (\ref{dispersion})
the speeds of sound and most of the features are reproduced with very good accuracy. 
It is well-known that in order to reproduce the flat feature in the TA modes near 
the X point, one must go well beyond the fifth neighbor.
For the band structure and the density of states (DOS), the overall agreement is good, 
except for the Gruneisen parameters of the TA branch,
where our calculations, which only include cubic force constants up to the first neighbor shell,
overestimate $\gamma(X,$TA). Based on Klemens' formula (Eq. (\ref{klemens})), one might
anticipate that our model will slightly underestimate the lifetime of TA modes
and thus their contribution in the thermal conductivity.

\subsection{Phonon lifetimes and mean-free paths}

\begin{figure}[t]
\includegraphics[angle=270,width=0.5\textwidth]{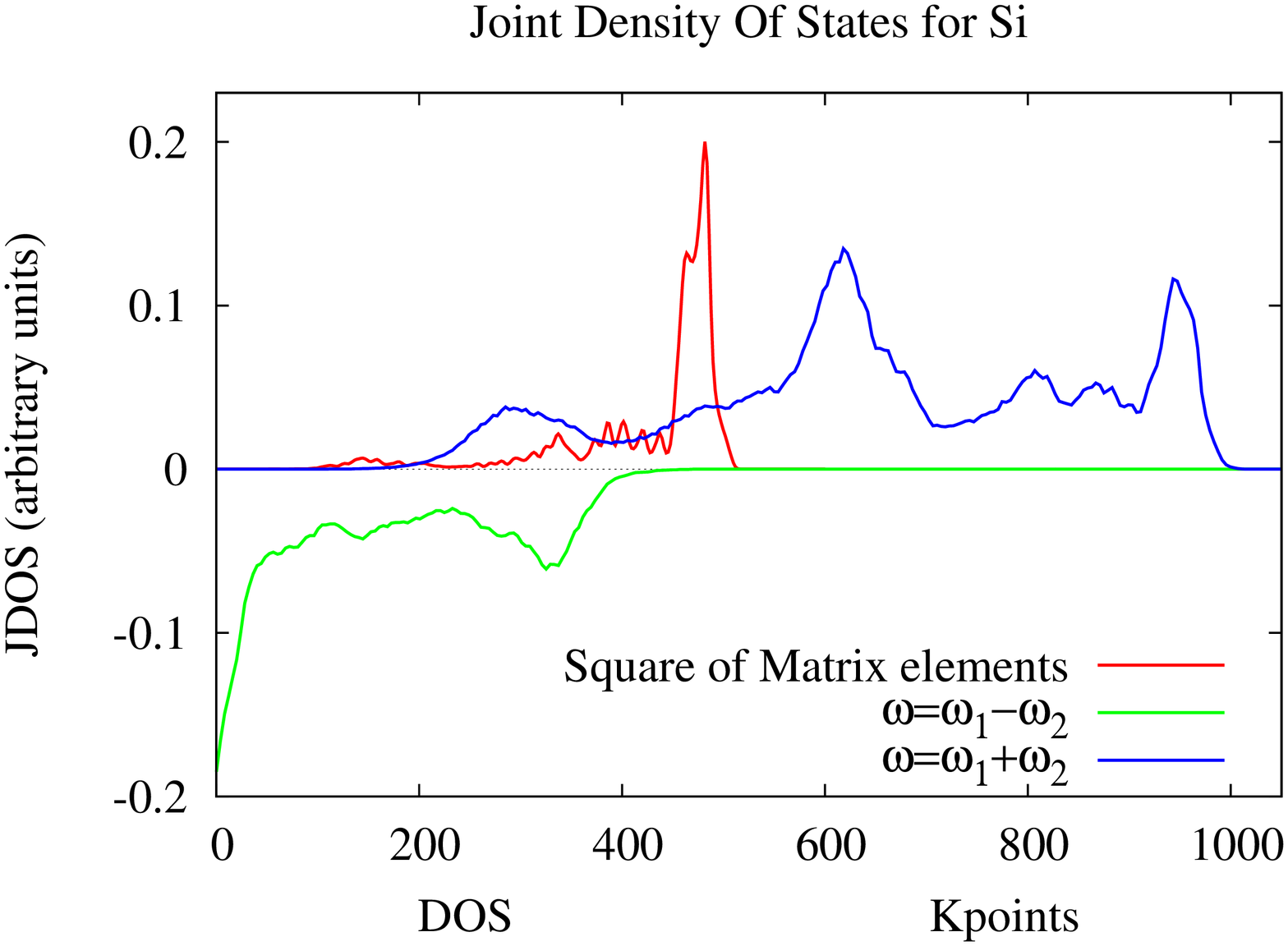}
\caption{(Color online) Top in blue is the DOS associated with two-phonon creation or annihilation
 ($DOS_2^-$),
and bottom in green is the DOS associated with one phonon emission or absorption ($DOS_2^+$).
In red, the contribution of the matrix elements defined in Eq. (\ref{3mtrx}) are displayed.
The peak at 500 cm$^{-1}$ is the main reason for smaller lifetimes of optical modes.
} 
\label{jdos} 
\end{figure}

To get an idea about the relative contributions of the matrix elements,
representing the strength of the 3-phonon interactions, versus the phase
space available for these transitions, characterized by the
two-phonon DOS, we show in Fig. (\ref{jdos}) the plots of these quantities.
We define the contribution of the matrix elements as: 
\beq
F(\om)=\sum_{\kl} \, \delta(\om - \om_{\kl}) \, \sum_{1,2} \, |V(\kl,1,2)|^2
\label{3mtrx}
\eeq
From Fig. (\ref{jdos}) we can note that optical phonons have a much larger weight coming from the 
matrix element $|V(\kl,1,2)|^2$. This explains why they have such a larger 
relaxation rate compared to acoustic modes for which the matrix elements
contribution is very small. The two-phonon DOS is representative of the phase
space available for the transitions, and is defined as:
\beq
DOS_2^{\pm}(\om)=\sum_{1,2} \delta (\om-\om_1 \pm \om_2)
\eeq
From Fig. (\ref{jdos}) it can be inferred that one phonon
absorption or emission ($DOS_2^+$) dominates for low frequency phonons (acoustic), while
two-phonon absorption or emission ($DOS_2^-$) dominates at high frequencies (LA and optical).

\begin{figure}[h]
\includegraphics[width=0.50\textwidth,angle=0]{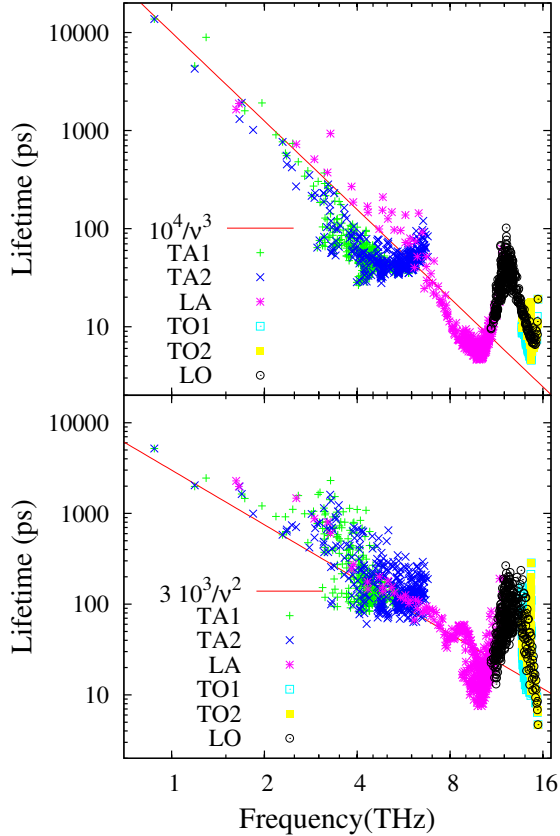}
\caption{(Color online) Lifetimes of the 6 branches in Si at 277 K versus frequency on a 
logarithmic scale. Top is for normal and bottom is for umklapp processes.
The quadratic dependence of the acoustic modes can be noticed for normal processes, 
while umklapp
processes seems to scale as $1/\om^3$. } 
\label{gama} 
\end{figure}

Next, we show in Fig. (\ref{gama}) the calculated lifetimes of the 3 acoustic 
and optical modes versus frequency for a regular mesh of kpoints in the first
Brillouin zone, at T=70 and 277K. The results depend slightly on the number of k-mesh 
points used for the integration within the FBZ. Here, we are showing
results obtained with 18x18x18 mesh, which is close to convergence.
The normal and umklapp components of the lifetimes are separated as $1/\tau = 1/\tau^U +1/\tau^N$.
We can note that although the lifetimes associated with normal processes are in $1/\om^2$,
those of umklapp processes seem to scale at low frequencies like $1/\om^3$ so
that the former dominates at low frequencies.
This is in contrast to the first-principles results provided by Ward and 
Broido\cite{ward} where they report that the umklapp rate is in $\om^4$. 
Even though not explicitly mentioned in their paper\cite{private}, fits to
their data with  $\om^3$ was almost as good as the fit with  $\om^4$.
In the appendix, we provide a proof why in the case of Si the umklapp
rate would behave as $\om^3$.

From Fig. (\ref{gama}), we can notice that at low frequencies (typically below 3 THz
or 100 cm$^{-1}$ where dispersions are linear),
normal rates dominate while at higher frequencies and typically for optical modes, 
umklapp processes dominate transport.

\subsection{Thermal conductivity from lattice dynamics}

\begin{figure}[h]
\includegraphics[angle=270,width=0.45\textwidth]{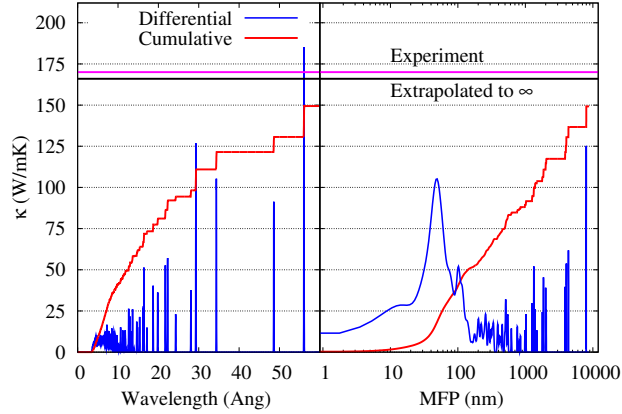}
\caption{(Color online) Cumulative contributions of phonons to the thermal conductivity at 277 K from
the 18x18x18 k-mesh data. Left is according to the wavelengths, and right is
according to the MFPs.
Both  differential and cumulative thermal conductivities are shown in blue and red respectively. 
For comparison, the extrapolated (to infinite k-mesh) and experimental $\kappa$
are also shown with horizontal lines at 166 and 174 W/mK respectively.
}
\label{mfp} 
\end{figure}

To see what is the contribution of each MFP to the total thermal conductivity, following
the approach of Dames and Chen\cite{dames},
we have decomposed the thermal conductivity based on each mode and sorted each component 
according to
their mean free paths. One can then define a differential thermal conductivity and 
the accumulated one, which is its integral:
\beqnar
d\kappa (\La_{\kl}) &=& {1 \over 3} v_{\kl}\,  \La_{\kl}\, {C_v}_{\kl} \nonumber \\
\kappa(\La) &=& {1 \over  N_k} \sum_{\kl}^{\La_{\kl} < \La} d\kappa (\La_{\kl})
\label{mfps}
\eeqnar
The above can be plotted versus the MFP, $\La$, seen as an independent variable.
Fig. (\ref{mfp}) shows such contribution at 277 K. Considering the extrapolated
value to be 166 W/mK, one can notice that MFPs
extend well beyond 10 microns  even at room temperature. MFPs longer than 1 micron
contribute almost to half of the total thermal conductivity!
One should also note that the range of MFPs in Si at least, span over 5 orders of 
magnitude from a nanometer to 100 microns at room temperature. This would be larger
as we go to lower temperatures.

\begin{figure}[t]
\includegraphics[angle=270,width=0.50\textwidth]{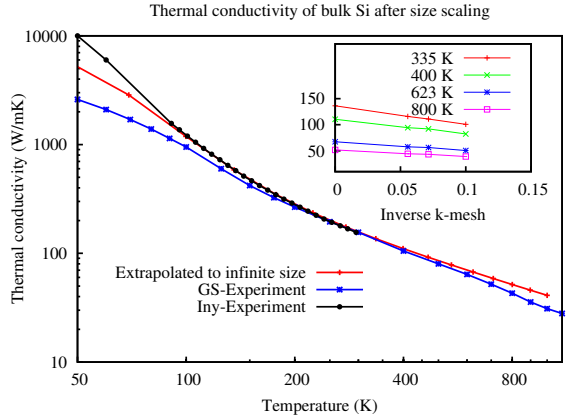} 
\caption{(Color online) Thermal conductivity of pure Si crystal from Eq. (\ref{rta}) versus temperature. 
Inset shows the extrapolation to infinitely dense kpoints.}
\label{kappa} 
\end{figure}

To get an acurate estimate of the thermal conductivity, one needs to 
extrapolate the data obtained from a finite number of k-mesh points, according
to Eq. (\ref{scaling}).
The extrapolated thermal conductivity versus temperature is plotted in Fig. (\ref{kappa})
and compared to the experimental results of Glassbrenner and Slack \cite{slack-si} and
Inyushkin {\it et al.}\cite{inyushkin}. We can notice that at low temperatures, boundary scattering
limits the experimental thermal conductivity. The agreement is very good in the temperature
range of 100 to 500K, after which experimental results decay faster due to higher order
phonon scatterings which are like $1/T^2$ or higher. 
Our resutls are within the relaxation time approximation, but one could also go
beyond and iteratively solve Boltzmann equation as Broido {\it et al.} have done\cite{broido08}.
They have shown that for Si and Ge, there would be about a further $10\%$ increase in $\kappa$. 

\begin{figure}[t]
\includegraphics[angle=270,width=0.5 \textwidth]{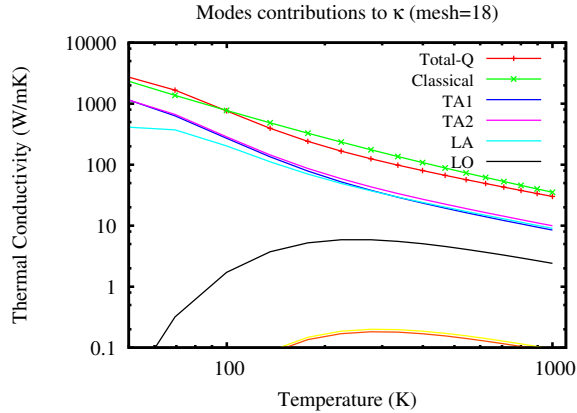}
\caption{(Color online) Quantum and classical thermal conductivities of Si 
versus temperature  (from Eq. (\ref{rta}) with k-mesh=18). 
The contribution of each mode is also being plotted.
}
\label{kappa-cl} 
\end{figure}

To assess the effect of the classical approximation, which is made in classical MD 
simulations, we have also compared in Fig. (\ref{kappa-cl}) for a given k-point density,
the classical and the quantum thermal conductivities within the RTA. They are displayed with 
symbols on the lines.
The quantum one is given by Eq.(\ref{rta}), and the classical one uses the same expression
in which the Bose-Einstein distribution is substituted by $k_BT/\hbar \om$ both in the 
heat capacity and in the relaxation time. We can notice that the difference is small
above the Debye temperature, as expected, but the classical value overestimates the
quantum one by 10 to $20 \%$ as the temperature is lowered further.
This is a combination of the larger heat capacity and a smaller lifetime in the classical case.
We have also plotted the contribution of each mode to the thermal conductivity.
We can note that at low temperatures maily the two TA modes equally contribute to $\kappa$,
whereas at temperatures above 200 K, LA and TA modes equally contribute about almost
1/3 of the thermal conductivity, while LO's contribution is about 5\%.

The computation of the thermal conductivity using the RTA is to some extent more straightforward
than the use of GK-MD. The former involves a double summation in the FBZ and has very
little systematic error in it, whereas the MD simulations require an ensemble averaging process with
a relatively large error bar, not to mention the much longer CPU time needed to run the MD simulations.

For a mesh of kpoints equal to the number of primitive cells included in the MD supercell, 
we have obtained agreement between MD results and the classical version of Eq.(\ref{rta}), as
also shown in table (\ref{kappa7-10}). 

\section{Conclusions}
Using first-principles calculations, we developed a classical force field which was used both
in a molecular dynamics simulation and in the calculation of anharmonic phonon lifetimes. Both
methods provided an estimate for the thermal conductivity of pure crystalline silicon. 
The results of these two methods agreed for the same system size in the case where
$\kappa_{LD}$ was evaluated in the classical limit. GK-MD is however much more time-consuming
and includes large statistical errors. Furthermore it does not provide much
information besides the way the integrated autocorrelation converges with simulation time.
Size effects were discussed and arguments were provided why equilibrium MD
simulations converged relatively fast with respect to the supercell size.
Lattice dynamics, on the other hand, proved to be faster, more accurate, and
contain more useful information.
The use of a linear extrapolation versus the inverse of the size
led to a surprizingly good agreement with experiments.
Such extrapolation is justified for relaxation rates which are 
quadratic in frequency at low frequencies. 
The decomposition of $\kappa$ into the contribution of different mean free paths
showed that in Si MFPs span over 5 orders of magnitude from 1 nm to 100 microns at
room temperature, where about half of the thermal conductivity comes from MFPs 
larger than 1 micron.

The developed potential has the advantage of being amenable to systematic improvement by 
including more neighbor shells at the cost of heavier calculations. 
The approach of using the FGR for the estimation of relaxation rates
and the RTA or an improved approximation to $\kappa$ by solving the linearized Boltzmann
equation, allows one to obtain a relatively accurate estimate of the thermal conductivity of 
an arbitrary bulk crystalline structure from a few force-displacement relations obtained  
using  first-principles calculations, without any fitting parameters. This method paves the
way for an accurate prediction of thermal properties of nanostructured or composite 
materials in a multiscale approach, which takes as input the relaxation times due
to anharmonicity and defect scatterings. 

\section{Acknowledgements}

The authors wish to acknowledge useful discussions with Junichiro Shiomi, 
Joseph Feldman, Peter Young, and Asegun Henry.
We  thank Nuo Yang for providing the SW force-displacement data 
used in Fig. (\ref{directions}).

This work was supported as part of the Solid-State Solar-Thermal Energy Conversion Center (S$^3$TEC), an Energy Frontier Research Center funded by the U.S. Department of Energy, Office of Science, Office of Basic Energy Sciences under Award Number DE-SC0001299.

\section{Appendix}

In this appendix, we show the frequency-dependence of the umklapp rates. 
According to Eq. \ref{self}, the relaxation rate is a product of the 3-phonon matrix
element $ |V(\ql,1,2)|^2 $, a combination of occupation factors, and delta 
functions reflecting the constraints of energy  conservation. 
We will separately discuss the frequency dependence of the matrix element
and the phase space term. 

First, the sum over the second momentum 2 is cancelled by the constraint of 
momentum conservation, so that the relaxation rate is just the 3D integral over $q_1$ in
the FBZ. One of the dimensions can be integrated over by using the
identity 
\beqnar
\int d^3 q_1 \, \delta(\om+\om_{\qo}-\om_{\qt}) f(\qo) &=& \nonumber \\
\int d^3 q_1 \, \delta(q_1-q_o)/|v_{\qo}-v_{q+q_1 \lambda_2}|  f(\qo)  &=& \nonumber \\
\int d^2 S_{q_o} 1/|v_{q_o \lambda_1}-v_{q+q_o \lambda_2}| f(q_o \lambda_1 )  
\label{surface}
\eeqnar
where $q_o$ is the solution of $\om_{q\la}+\om_{q_o \la_1}-\om_{-q-q_o \la_2}=0$.
Note that the denominator containing the group velocities is not small 
as $\la_1$ as long as $\la_2$ refer to two different branches; but in case 
$\la_1=\la_2$, the denominator becomes linear in $q$.

Second, for umklapp processes, in the small $\om$ limit, we must have both $q_1$ and $q_2=-q-q_1$
near the Brillouin zone boundary such that $q_1$ is inside the zone and $q_2$ 
outside; so that the corresponding frequencies are not infinitesimally small, 
but their difference would be.
In general, this forces the $q_1$ surface integral to be limited 
to a pocket of dimensions $q$ located
at the FBZ boundary, so that the surface integral is of the order of $q^2$. 
But in case where there is a degenerate band at the zone boundary, 
the surface would be of order $q$ instead. 
Different cases based on the symmetry of the crystal and the 
type of degeneracy have been discussed in detail by Herring\cite{herring}.
In our case of interest, namely Si, it is possible to have 
a 3-phonon process involving a small momentum $q$ acoustic mode connecting 
the LA branch to the LO one, with which it is degenerate, near the Brillouin 
zone boundary all along $X \to W$,
with a surface area $S_{q_o}$, therefore, of order $q$.

Third, among the two types of terms: phonon $\om$ decaying to $\om_1+\om_2$ and 
one phonon absorption $\om+\om_1=\om_2$, the former cannot contribute because
$\om \simeq 0$ and $\om_1 $ and $\om_2$ are finite. Therefore only the terms
$ (n_2-n_1) \times [\delta( \om_1 -\om_2 + \om) -\delta( \om_1 -\om_2 -  \om) ]$
contribute to the umklapp lifetimes at small frequencies. In the latter, one
can substitute $n_1-n_2$ by $\pm \,\om \, \partial n/\partial \om_1 \simeq O(q) $. 
We must remember to substitute the argument $\om$ in the relaxation rate by 
its on-shell value $\om_q=v \times q \to 0$.    
So that, in the limit of low frequencies, the inverse lifetime can be written as:
\beq
\int d^2 S(q_o) \, {1 \over |v_{q_o \lambda_1}-v_{q+q_o \lambda_2}| } \, 
\om_{q\la} {\partial n \over \partial \om_o}  |V(q,q_o,-q-q_o)|^2
\eeq

Finally, due to the odd parity of the cubic force constants, one can show that for small
$q$ we have $ |V(q,q_o,-q-q_o)| \propto {\rm Sin} \,qR/\sqrt {\om_q} \propto {\sqrt q}$.

Putting everything together, we find that the umklapp rates at low
frequencies are, to leading order, of the form:
\beq
{1 \over \tau^U(\om)} \propto  q^3  \propto \om^3
\eeq
This is in agreement with our numerical findings. 

For normal processes, there is no restriction for modes 1 and 2 to be near the BZ
boundary. For instance, in the  (LA $\to$ LA + TA) process the term $(1+n_1+n_2)$ 
contributes and will not be linear in $q$. In such cases the rate would be in
$q^2$ and would dominate terms with higher powers of $q$.



\begin{references}


\bibitem{sw}
F. H. Stillinger and T. A. Weber,   Phys. Rev. B {\bf 31}, 5262 (1985). 

\bibitem{atb}
G. C. Abell, Phys. Rev. B {\bf 31}, 6184 (1985);
J. Tersoff, Phys. Rev. B {\bf 38}, 9902 (1988); 
D. W. Brenner, Phys. Rev. B {\bf 42}, 9458 (1990).

\bibitem{bo}
M. C. Payne, M. P. Teter, D. C. Allan, T. A. Arias, and J. D. Joannopoulos,
Rev. Mod. Phys {\bf 64}, 1045 (1992).

\bibitem{cp}
R. Car and M. Parrinello, Phys. Rev. Lett. {\bf 55}, 2471 (1985). 

\bibitem{green}
M. S. Green, 
J. Chem. Phys {\bf 22}, 398 (1954).  

\bibitem{kubo}
R. Kubo, 
J. Phys. Soc. Jpn. {\bf 12}, 570 (1957).  


\bibitem{magic} One reason for the success of the SW potential compared to others, can be 
attributed to its ability to reproduce a relatively correct Gruneisen parameter and thermal
expansion coefficient. It is not clear to us whether or not this is accidental, as the way 
SW determined their parameters was not clarified in their paper. The properties of SW and Tersoff
potential have been discussed by Porter, Justo and Yip in Jour. Appl. Phys. {\bf 82}, 5381 (1997).

\bibitem{volz}
S. G. Volz and G. Chen,
Phys. Rev. B {\bf 61}, 2651 (2000).

\bibitem{philpot}
P. K. Schelling, S. R. Phillpot, and P. Keblinski
Phys. Rev. B {\bf 65}, 144306 (2002). 

\bibitem{ase}
A. Henry and G. Chen, 
Jour. Comput.  Theor. Nanosci., {\bf  5}, 141, 2008.

\bibitem{sellan}
D. P. Sellan, E. S. Landry, J. E. Turney, A. J. H. McGaughey, and C. H. Amon
Phys. Rev. B {\bf 81}, 214305 (2010). 


\bibitem{broido05}
D. A. Broido, A. Ward, and N. Mingo,
Phys. Rev. B {\bf 72}, 014308 (2005).

\bibitem{murthy1}
Sun and J. Murthy, APL {\bf 89}, 171919 (2006).

\bibitem{murthy2}
J. A. Pascual-Gutierrez, J. Murthy and R. Viskanta, JAP  {\bf 106}, 063532 (2010).

\bibitem{70}
Private communication with the referee.

\bibitem{broido08}
D. A. Broido, M. Malorny, G. Birner, N. Mingo, and D. A. Stewart, Appl. Phys. Lett. {\bf 91}, 231922 (2007).

\bibitem{keivan-stokes}
K. Esfarjani and H. T. Stokes,
Phys. Rev. B {\bf 77}, 144112 (2008). 

\bibitem{qe}
Quantum Espresso is an electronic structure package based on the density functional theory
developed at SISSA. The methodology is detailed in: 
P. Giannozzi {\it et al.}, JPCM {\bf 21}, 395502 (2009).


\bibitem{mingo}
N. Mingo, K. Esfarjani,  D. A. Broido, and D. A. Stewart,
Phys. Rev. B {\bf 81}, 045408 (2010).


\bibitem{statbook}
Time Series: Modeling, Computation, and Inference (Chapman \& Hall/CRC Texts 
in Statistical Science)

\bibitem{turney}
J. E. Turney, E. S. Landry, A. J. H. McGaughey, and C. H. Amon,
Phys. Rev. B {\bf 79}, 064301 (2009).

\bibitem{pz}
J. P. Perdew and A.
Phys. Rev. B {\bf 23}, 5048 (1981).

\bibitem{nilsson}
G. Nelin and G. Nilsson, Phys. Rev. B {\bf 5}, 3151 (1972).

\bibitem{marad}
A. A. Maradudin and A. E. Fein, Phys. Rev. {\bf 128}, 2589 (1962).

\bibitem{cowley}
R. A. Cowley, Rep. Prog. Phys. {\bf 31}, 123 (1968).

\bibitem{srivastava}
G. P. Srivastava, ``The physics of phonons'', Taylor and Francis (1990).

\bibitem{reissland}
J. A. Reissland, ``The physics of phonons'', J. Wiley (1973).

\bibitem{slack-si}
C. J. Glassbrenner and G. A. Slack,
Phys. Rev. {\bf 134}, A1058 (1964).

\bibitem{inyushkin}
A. V. Inyushkin,  A. N. Taldenkov, A. M. Gibin, A. V. Gusev, and H.-J. Pohl,
Phys. Stat. Sol. (c) {\bf 11}, 2995 (2004).

\bibitem{klemens}
P. G. Klemens, In Thermal Conductivity (Edited by
R. P. Tve). {\bf 1} D. 1, Academic Press. London (1969).

\bibitem{ward}
A. Ward and D. A. Broido,
Phys. Rev. B {\bf 81}, 085205 (2010).

\bibitem{private}
After private communication with D. Broido, we found out that their low-frequency data
could as well be fitted with $\om^3$, and their choice of $\om^4$ was because
of a better agreement in some intermediate frequency range.

\bibitem{dames}
C. Dames and G. Chen, "Thermal Conductivity of Nanostructured Thermoelectric Materials," CRC Handbook, Ed. M. Rowe, Taylor and Francis, Boca Raton, (2006).

\bibitem{herring}
C. Herring, Phys. Rev. {\bf 95 }, 954 (1954).

\end{references}
\end{document}